

      \def\d{\partial}
      \def\D{{\cal D}}
      \def\ket#1{|\, #1\, \rangle}
      
      \def\loint{\oint\nolimits}
      \def\real{{\rm I\!R}}
      \def\to{\rightarrow}
      \def\tr{\mathop{\rm tr}}

\centerline{\bf Recent Developments in Classical and Quantum Theories of
Connections}
\centerline{\bf Including General Relativity}
\bigskip
\centerline{Abhay Ashtekar}
\centerline{Physics Department, Syracuse University, Syracuse, NY
13244-1130, USA}
\bigskip
\bigskip

General relativity can be recast as a theory of connections by
performing a canonical transformation on its phase space. In this
form, its (kinematical) structure is closely related to that of
Yang-Mills theory and topological field theories. Over the past few
years, a variety of techniques have been developed to quantize all
these theories non-perturbatively. These developments are summarized
with special emphasis on loop space methods and their applications
to quantum gravity.
\bigskip
\bigskip

{\bf 1. Introduction}

In the first conference in this series, held at Goa in 1988, I
presented some results on a new Hamiltonian formulation of general
relativity and outlined how these results could be used in the
construction of a non-perturbative quantum theory of gravity.
During the last three years, this general program has been pursued
vigorously by a number of groups. The key idea in the initial stage
was to exploit the fact that, in its new version, the phase space of
general relativity is the same as that of theories of connections
--gauge theories and topological field theories-- to import into quantum
gravity ideas and techniques from these better understood theories. It
soon turned out, however, that the methods initially developed for
quantum gravity could in turn be applied to other theories of connections.
Thus, there has been a synergetic exchange of ideas between general
relativity and gauge theories and a variety of new results have been
obtained. Indeed, in the space allotted to me it would be impossible
to do justice to all the developments that have occurred. Fortunately,
there already exist in the literature a number of detailed reviews:
there is a monograph [1] addressed to research students which treats all
the basic issues with due care to mathematical subtleties; there is a
more compact review [2] which appeared in a journal, addressed to standard
physics audiences; and, there is a summer school report [3], geared to
particle physicists and field theorists, which emphasizes the more recent
work in this area.

In this article, therefore, I shall not attempt to present a
comprehensive summary. Rather, my aim is to outline the main directions
in which the work has proceeded since the Goa conference and to supply
references where details can be found. When one tries to squeeze diverse
themes along which research has naturally progressed into a few
categories, there are inevitable omissions. I apologize in advance to
my colleagues if they find that their favorite ideas have unfortunately
been skipped.

I have divided the material according to themes. Consequently,
individual sections vary a great deal in their length.
Section 2 is devoted to the developments in classical general relativity
and geometry. Section 3 addresses the general quantization program,
emphasizing the points at which it goes beyond Dirac's treatment of
quantization of constrained systems. Section 4 is devoted to the
quantum theory. Open problems and directions of current work are
summarized in section 5.
\bigskip
{\bf 2. Hamiltonian general relativity and geometry:}

In the new Hamiltonian framework, it is convenient to begin with complex
general relativity. The phase space then is the same as that of $SO(3)$
Yang-Mills (or, topological field) theory:
the configuration variable is a complex-valued, $SO(3)$ connection,
$A_a^i(x)$, and the canonically conjugate momentum is a vector density
${E}^a_i(x)$ of weight 1, where $a$ is a (tangent space) vector index
while $i$ is an internal, $SO(3)$ index. In Yang-Mills theory, ${E}^a_i$
has the interpretation of the electric field. In general relativity, it
represents a (density weighted) triad. There are seven first class
constraints:
$$\D_a{E}^a_i = 0;\quad {E}^a_i\- F_{ab}^i = 0; \quad {\rm and}\quad
\epsilon^{ijk} E^a_iE^b_j\- F_{abk} = 0; \eqno(1)$$
where $\D$ is the gauge covariant derivative operator defined by
$A_a^i$ and $F_{ab}{}^k := 2\d_{[a}A_{b]}{}^i +G \epsilon^{ijk} A_{aj}
A_{bk}$ is the field strength of $A_a^i$. (G is Newton's constant. Note
that $GA_a^i$ has dimensions of inverse length, the usual physical
dimensions of a connection.) Since the first three of
these constitute precisely the Gauss constraint of gauge theories,
we see that the constraint surface of general relativity is in fact
embedded in that of $SO(3)$-gauge theories. The remaining four
constraints are the ``genuine relativity constraints''. The first
three of these --called collectively the vector constraint-- generate
spatial diffeomorphisms while the last one --the scalar constraint--
generates ``pure'' time evolution. Apart from a surface term, the
Hamiltonian is a linear combination of these seven constraints. Like
the constraint functionals, the evolution equations are low order
polynomials in the basic canonical variables. For example, setting
the Lagrange multipliers of the Gauss and the vector constraints
equal to zero, for ``pure'' time evolution, we have the following
equations of motion:
$$\dot{A}_a^i = iNE^b_j F_{bak}\epsilon^{ijk} \quad {\rm and} \quad
  \dot{E}^a_i = i \D_b(NE^a_jE^b_k) \epsilon_i{}^{jk} \eqno(2)$$
where $N$ is the lapse, i.e., the Lagrange multiplier of the scalar
constraint. All these equations are significantly simpler than
those encountered in geometrodynamics, where the 3-metric, rather
than a connection, is the configuration variable. In addition, the
Hamiltonian description of general relativity is now closely related
to that of gauge theories. As we will see, these features have been
exploited in a number of ways in the recent developments.

Note however that so far we have considered complex relativity. To
recover the real theory, we have to restrict ourselves to the
appropriate ``real section'' of the complex phase space. In the early
papers there was some confusion on the expression and implementation
of the reality conditions. This issue has been clarified now: the
reality conditions again involve only low order polynomials in the
basic canonical variables. (See [4] and chapters 8, 3.3 and 4.3 of
[1]. While I will restrict myself to the source-free theory in this
article, all results mentioned above have been extended in [4,5] to
include matter fields and the cosmological constant.)
In Euclidean general relativity, these conditions require simply that
$A_a^i$ and $E^a_i$ be real. In the Lorentzian theory, on the other
hand, although they involve expressions which are (at worst) only quartic
in $A_a^i$ and $E^a_i$, their implementation is not so straightforward.
Thus, it does appear that the simplicity of the field equations has
been attained, to some extent, at the cost of having to deal with the
reality conditions. For example, the dynamical trajectories in the space
of connections can be interpreted as ``null geodesics'' of a super-metric.
This interpretation provides a useful technique in Bianchi models to find
the qualitative behavior of solutions and, in many cases, complete
solutions. However, in general, it is hard to pick out the null
geodesics which would correspond to real, Lorentzian solutions. More
generally, the implementation of the Lorentzian reality conditions remains
one of the most important open problems in the program. However, some work
{\it has been} done to make these conditions more manageable and I will
return to this point in sections 3-5.

Since the basic variable is a connection, it is tempting to look for
phase space variables which are manifestly gauge invariant and base
quantization on them. For configuration variables, there is an obvious,
natural choice: traces of holonomies around closed loops. More precisely,
given any closed loop $\gamma$ (in the ``spatial'' 3-manifold on which
$A_a^i$ are defined), we can set:
$$Q_\gamma (A) := {1\over 2} \tr{\cal P}\> G\loint_\gamma dl^a A_a\> ,
\eqno(3)$$
where trace is taken in the spin $\textstyle{1\over 2}$-representation of
$SO(3)$, ${\cal P}$ stands for ``path-ordered'' and $G$, as before, is
Newton's constant. The vector
space generated by these $Q_\gamma$ is closed under the product because
of the Madelstam identities [6,7]. To get momentum variables --i.e.,
functions on the phase space which are linear in momenta-- we introduce
2-dimensional (closed) strips. A strip $S$ is a mapping from $S^1\times
(-\epsilon, \epsilon)$ to the 3-manifold. Thus, a strip is coordinatized
by two parameters, $\sigma \in [0, 2\pi]$ and $\tau\in (-\epsilon,
\epsilon)$. (These coordinates can be changed by a reparametrization. The
invariant structure is the topology $S^1\times\real$ and the foliation of $S$
by a fixed family of circles.) With each strip $S$, we associate a
momentum variable
$$P_S (A,E):= \int_S dS^{ab} \eta_{abc} \- \tr E^c\-
    U_\tau(\sigma, \tau ), \eqno(4)$$
where $U_\tau$ is the group element representing the (untraced) holonomy
of the connection around the loop $\tau = {\rm const}$. Clearly, these
functions are also gauge invariant. Now, the surprising result is that
the loop-strip (i.e these configuration and momentum) variables are in
fact closed under the Poisson bracket! Furthermore, the structure
constants involved can be constructed from simple geometric properties of
loops and strips in three-dimensional space [1,2, 7]. Finally, these
variables are (over)complete almost everywhere on the phase space
\footnote{An exhaustive characterization of regions (sets of
``measure zero'') of the phase space where they fail to be complete and the
structure of the intersection of these regions with the constraint surface
and the real section of the phase space is available. However, results of
[8] are based on the ``small $T$-algebra'' of Rovelli and Smolin [6] which
is closely related to but not the same as the loop-strip algebra discussed
above. Therefore, strictly speaking, some further work is needed to show
that the last set of results holds also for the loop-strip algebra.}[8].
It is therefore natural to use the loop-strip algebra as the starting point
in quantum theory.

The simple form of the constraints has led to two ingenious methods of
solving them. Capovilla, Jacobson and Dell (CDJ) [9] have pointed out that
the scalar and the vector constraint can be solved simply by making an
{\it algebraic} ansatz for $E^a_i$ in terms of $A_a^i$. Furthermore, the
solution is generic. This is a striking result. However, one is still left
with the Gauss constraint and a useful technique to impose it on the CDJ
``free data'' has not yet emerged. More recently, Newman and Rovelli [10]
have used Hamilton-Jacobi techniques to address a more important problem:
that of finding {\it gauge invariant} free data. They have been able to
treat in this way the Gauss and the vector constraints. Thus their analysis
provides elegant geometric coordinates on the phase space reduced with
respect to these six constraints. Work is in progress on the scalar
constraint. If this program is carried to completion, one would have
available a complete set of Dirac-observables for full general relativity.
These would play a key role in quantization. Note, however, that both the
CDJ and the Newman-Rovelli techniques have so far been applied to the
{\it complex} theory and imposition of reality conditions does not appear
straightforward.

The choice of connections $A_a^i$ as the configuration variables is also
natural from the viewpoint of dynamics of the theory. First, there is
a striking geometric result [8]: the holonomy group of $A_a^i$ is
the same on any Cauchy slice of the real, Lorentzian space-time; it is a
``constant of motion''. Second, because the scalar constraint is purely
quadratic in momenta --in contrast to geometrodynamics, there is no
``super-potential'' in connection-dynamics-- the dynamical trajectories
in the connection superspace can be interpreted as the null geodesics of
the ``super-metric'' $\epsilon^{ijk}F_{abk}$. There are thus numerous
advantages in regarding general relativity as a theory of dynamics of
(self-dual) 3-connections rather than of 3-metrics. This viewpoint was
pushed to its logical extreme by Capovilla, Dell and Jacobson [9]. They
begin with an action which is built out only of self-dual connections and
which, in its Hamiltonian version, reproduces the description given above.
In this description then, the ``triads'' $E^a_i$ appear only as momenta
canonically conjugate to the 3-connection. The action, or indeed the
entire framework, knows nothing about a space-time metric which can, if one
wishes, be introduced as a secondary field constructed from the curvature
of the connection. This analysis has had a number of interesting off-shoots.
For example, it led Capovilla [11] to construct a family of ``neighbors
of general relativity''. In the Hamiltonian version of these theories, the
configuration variable is again a connection, the Gauss and the vector
constraints have the same form as in Eq. (1) but the scalar constraint has
additional terms. Recently, Peld\'an [12] has taken this work further by
showing that one can replace the internal group $SO(3)$ by any Lie group
of ${\rm dim} \ge 3$; the theory then represents gravity coupled to matter
although the physical meaning of these couplings is still unclear.

Finally, there are the applications of the framework to geometry. A number
of interesting results about half-flat metrics as well as Einstein spaces
with self-dual Weyl curvature have been obtained in the Riemannian (i.e.,
++++) regime [13]. Perhaps the most striking of these is the exhaustive
analysis of the structure of the moduli spaces of gravitational instantons
with self dual Weyl curvature and positive cosmological constant
due to Torre [14]. This imaginative analysis is an illustration of how
the relation between the Yang-Mills theory and general relativity
discussed above can be exploited to obtain rigorous results of interest
to differential geometry.
\bigskip

{\bf 3.Extension of the Dirac program}

Attempts at canonical quantization of general relativity have
traditionally followed the general program introduced by Dirac for
quantization of constrained systems [15]. The Dirac program, however,
is incomplete in one important respect: while it tells us that
constraints should be incorporated in the quantum theory as conditions
which select the physical states, it provides no guidelines for
introducing the appropriate inner product on the space of these
physical states. In particle mechanics or Minkowskian field theories
this does not pose problems in practice because one can use the
available symmetries to select a preferred inner-product. For examples,
in Minkowskian field theories, one selects the vacuum state by invoking
Poincar\'e invariance and then uses the vacuum expectation values of
physical operators to obtain the required Hilbert space structure. In
quantum gravity, on the other hand, such a space-time group of symmetries
is simply not available. (One might imagine using the spatial
diffeomorphism group to select the vacuum. However, the strategy fails
because {\it every} physical state is invariant under this group.)
Therefore, Dirac's program has to be supplemented with a new guiding
principle.

While this problem exists for {\it any} canonical approach to quantum
general relativity, there are additional features which are peculiar to
our specific approach which also require an extension of the Dirac
program. First, for real Lorentzian relativity, the ``canonical''
variables constitute a hybrid pair: the Lorentzian reality conditions
require that $E^a_i$ be real, while $A_a^i$ is allowed to be complex (but
such that the time-derivative of $E^a_i$ is again real).
Consequently, the canonical quantization procedure itself
is at first somewhat obscure. For example, in the $A_a^i$ representation,
the states $\Psi(A)$ are now {\it holomorphic} functionals of connections
$A_a^i$. Can one still represent the $E^a_i$ operator by the functional
derivative with respect to $A_a^i$? A second problem is associated with
the loop-strip variables discussed in section 2. Because they are
manifestly gauge invariant and closed under the Poisson brackets, it is
attractive to use their Poisson algebra as the starting point in
canonical quantization. However, unlike the habitual canonical pairs
such as the metric and the extrinsic curvature, the loop-strip variables
are {\it over}complete almost everywhere on the phase space. How does one
handle this overcompleteness in the quantum theory? Further, there {\it
are} regions (although of ``measure zero'') of the classical phase space,
where these variables fail to be complete. Do we now have a problem just
of the opposite sort in quantum theory?

Since the Goa conference, these issues have been analysed in detail and
an appropriate extension of the Dirac program has been constructed.
(See chapter 10 of [1].) The idea is to use an algebraic approach.
One first constructs an abstract algebra of quantum operators based on a
given Poisson algebra, and then looks for its representations. The
possible relations that may exist due to overcompleteness of the basic
classical variables --such as the loop-strip functionals-- are now built
in to the very structure of the quantum algebra. Similarly, the reality
conditions in the classical theory are to be coded in the
$\star$-relations of the quantum algebra. These $\star$-relations in turn
are to determine the inner-product: one seeks that inner-product with
respect to which the $\star$-relations, defined abstractly on the algebra,
are realized as the Hermitian adjoint relations on the Hilbert space.
Thus, far from being a nuisance, the classical reality conditions may
ultimately determine the quantum inner-product. This specific idea
has been tested in source-free Maxwell theory [16,17], linearized gravity
[18] and 2+1-dimensional gravity ([19] and chapter 17 of [1]). The first
two examples are especially instructive because in these Minkowskian
field theories, the new strategy leads one to the correct inner product
{\it without} having to appeal to the Poincar\'e group. Finally, in this
program, the issue of time is decoupled from that of finding the
inner-product. In the 2+1-theory, for example, the issue of singling out
a time variable is almost as hard as in the 3+1 theory; both theories are
diffeomorphism invariant and therefore devoid of a background metric.
Using the general program sketched above, we are able to find the
required inner product on physical states {\it without having to}
first isolate the time variable.

This extension of the Dirac program is meant to provide general guidelines;
it is not a rigid set of rules. It has been carried out to completion in
a number of examples which mimic various features of the new Hamiltonian
formulation of general relativity [20] and this detailed analysis has
provided confidence in the underlying ideas. It has also given rise to
a number of conjectures. For example, we suspect that if there exists an
inner-product which can implement the $\star$-relations faithfully and if
the resulting representation of the algebra is irreducible, the inner
product is unique (up to an overall constant factor). Such conjectures
remain unproven however and further work is clearly needed. Much of this
work would rely only on general techniques from mathematical physics; these
results should therefore have validity well beyond quantum general
relativity.
\bigskip

{\bf 4. Quantum general relativity}

In quantum general relativity, several developments have occurred. The
results obtained so far clarify a number of issues and provide considerable
support for the strategies which were outlined at Goa. Furthermore, as we
shall see, some of these results are quite striking and extremely
encouraging. Nonetheless, the program as a whole is still far from being
complete. For example, while a number of new solutions to all quantum
constraints have been found, the set of solutions obtained so far is
still quite incomplete. Perhaps more importantly, there remain several
conceptual problems. Some of these are common to any non-perturbative
approach to quantum gravity. Examples are: singling out useful observables,
resolving the issue of time and interpreting the framework as a whole.
There are others which are specific to the approach being pursued. By
now, there does exist a large body of results that has been obtained using
the loop variables and the general quantization techniques summarized in
section 3. However, when looked at in detail, one finds that these results
do not quite fit together in to a precise and uniform mathematical
framework.
There are several gaps that need to be filled before we have ``global
understanding''. (One should perhaps emphasize, however, that this issue
has become relevant precisely because the program has reached a certain
level of maturity, not enjoyed by other non-perturbative approaches to
quantum general relativity.) In this section, I will summarize the results
obtained so far as well as the puzzles that still remain.
\medskip

{\sl 4.1 Regularization and Weaves}

As explained in the Goa conference, Rovelli and Smolin [6] introduced a
new representation for quantum general relativity in which states are
certain functionals of closed loops. This provides a representation of
the loop-strip algebra on a vector space --the space of loop states
\footnote{One {\it can} equip the space of loop states with an
Hermitian inner product (using the Gel'fand-Naimark-Segal construction)
in which, moreover, the diffeomorphism group of the underlying 3-manifold
acts unitarily. However, it is not obvious that this is the ``correct''
inner-product especially because the resulting Hilbert space is
non-separable. Therefore, most results have been obtained using just the
vector space structure of the space of loop functionals.}.
The action of the loop-strip operators involves simple geometric
operations such as breaking, re-routing and gluing loops. This is the
starting point for quantization. Perhaps the most interesting of the
``kinematic'' results so obtained are the following: Regularization of
operators respecting the (3-dimensional) diffeomorphism invariance
of the theory and existence of states that approximate a flat metric
on a large scale but necessarily exhibit a discrete structure on the
Planck scale. These states are called {\it weaves}; they represent how
a macroscopic, classical geometry can be ``woven'' using excitations along
loops as ``quantum threads''.

Let us begin with regularization.
Since a basic phase space variable is a (density weight one) triad
$E^a_i(x)$, the spatial metric (of density weight two) is a ``composite''
field given by $q^{ab}(x) = E^a_i(x) E^b_i(x)$. In the quantum theory,
therefore, this operator must be regulated. The obvious possibility is
point splitting. One might set $q^{ab}(x) = \lim_{y\to x} E^{ai}(x)
E^b_i(y)$. However, the procedure violates gauge invariance since the
internal indices at two {\it different} points have been contracted. A
gauge invariant prescription is to use the Rovelli-Smolin loop variable
$T^{ab} [\gamma ](x,y)$ defined in the classical theory by
$$T^{aa'}[\gamma ] (y,y') := \tr\big[({\cal P} \exp\- G\int_{y'}^y A_a
dl^a) E^a(y')\- ({\cal P}\exp \- G\int_y^{y'}A_a dl^a)\- E^{a'}(y)\big],
\eqno(5)$$
where $y$ and $y'$ are any two points on the loop $\gamma$, and note that
in the limit $\gamma$ shrinks to zero, $T^{aa'}[\gamma](y,y')$ tends to
$q^{aa'}$. In quantum theory, one can formally define the action of the
operator $\hat{T}^{aa'}[\gamma](y,y')$ directly on the loop states.
As with other loop operators, it acts by breaking and re-routing the loops
that appear in the argument of the quantum state. One may therefore try
to define a quantum operator $\hat{q}^{aa'}$ as a limit of $\hat{T}^{aa'}
[\gamma ]$ as $\gamma$ shrinks to zero. The resulting operator does exist
after suitable regularization and renormalization. However, because of the
density weights involved, the operator necessarily carries memory of the
background metric used in regularization. Roughly, this comes about as
follows. The operator in question is analogous to the product
$\delta^3(x)\cdot\delta^3 (x)$ of distributions at the same point. To
regulate it, we have to introduce a background metric. The final result is
a distribution of the form $N(x)\delta^3(x)$ where, because $\delta^3(x)$
is a density of weight one, the renormalization parameter $N$ is now a
density of weight one, proportional to the determinant of the background
metric. Since the final answer carries a memory of the particular
metric used to regulate the operator, we have violated diffeomorphism
invariance. Although there is no definitive proof, there do exist arguments
which suggest that any {\it local} operator carrying the information about
geometry will face the same problem.

There do exist, however, {\it non-local} operators which can be regulated
in a way that respects diffeomorphism invariance [21,3]. Furthermore, the
resulting operators are finite {\it without} the need of any
renormalization. Thus, there are no free renormalization constants in
the final expressions. As the first example, consider the function
$q(\omega)$ --representing the smeared 3-metric-- on the classical
phase space, defined by
$$q(\omega):= \int d^3x\> (q^{ab}\omega_a\omega_b)^{1\over 2}\- ,
\eqno(6)$$
where $\omega_a$ is any smooth 3-form of compact support. (Note that the
integral is well-defined without the need of a background volume element
because $q^{ab}$ is a density of weight two.) The corresponding operator
is defined as follows: First re-express $q(\omega)$ using the loop
variable $T^{ab}$ of Eq.(5), then replace $T^{ab}$ by a regulated operator
$\hat{T}^{ab}_\epsilon$ on the loop states and finally take the limit
as the regulator $\epsilon$ goes to zero. The result, $\hat{q}(\omega )$,
is a well-defined operator on the loop states; it carries no imprint of
background structures used in regularization. For example, if $\gamma$
is a smooth loop without self-intersections, we have a rather simple
action:
$$\hat{q}(\omega )\circ \Psi (\gamma ) = \l_P^2 \oint_\gamma ds
|\dot{\gamma}^a \omega_a|\>\cdot \Psi(\gamma) ,\eqno(7)$$
where $l_P= \sqrt{G\hbar}$ is the Planck length, $s$, a parameter along
the loop and $\dot{\gamma}^a$ the tangent vector to the loop. (The $G$
in $l_P$ comes from the fact that $GA_a^i$ has the usual dimensions of
a connection (see Eq.(3)) and $\hbar$ comes from the fact that
$\hat{E}^a_i$ is $\hbar$ times a functional derivative.) One can similarly
define the operator corresponding to the area $\hat{\cal A}_S$ of a smooth
2-surface $S$. The result is:
$$\hat{\cal A}_S \circ \Psi (\gamma ) = l_P^2 \>I(\gamma ,S)\cdot\Psi
(\gamma )\> ,\eqno(8)$$
where $I(\gamma, S)$ is the unoriented intersection number between the
loop $\gamma$ and the surface $S$. Thus, the final result is simple
and geometrical: each intersection of the loop with the surface contributes
a Planck area to the surface. If the state has a support on a single loop,
it would clearly {\it not} resemble a classical geometry; most surfaces
would have no intersection with that loop whence the area of most surfaces
would be simply zero. Such a state would represent ``an elementary''
excitation of geometry; like a 1-photon state in the Maxwell theory, it
wold not have a classical analog. To approximate a classical geometry,
the state must involve many loops so that given any smooth surface $S$
with area $A$ in the classical geometry, there are approximately
$A/l_P^2$ intersections between $S$ and the loops.

With these operators at hand, we can now introduce weaves. Fix on
$\real^3$ a flat metric $h_{ab}$ and let us ask if we can construct
states which approximate $h_{ab}$ on a scale $L$ which large compared to
$l_P$.
\footnote{Note the logic of the argument: we first fix a flat metric
we want to approximate and then use it repeatedly to define the length
scales need in the argument. This procedure is necessary because in
non-perturbative quantum gravity there is no background metric and
therefore no a {\it physical} notion of length. In particular,
although we can set $\l_P = \sqrt{G\hbar}$, this quantity does not
represent a physical length unless we have a metric which enables one
to measure lengths.}
This question can be phrased without reference to a specific inner-product
because the operators $\hat{q}(\omega)$ and $\hat{\cal A}_S$ carrying
information about geometry are ``diagonal'' in the loop representation.
That is, using the form of the operators (7) and (8) we can ask:
Are there states $\Psi(\gamma)$ for which
$$ \hat{q}(\omega) \circ \Psi (\gamma) = [h(\omega ) + O({l_P\over L})]
\cdot \Psi[\gamma ],\eqno(9)$$
where $h(\omega) := \int d^3x \- (h^{ab}\omega_a\omega_b)^{1\over 2}$
is the value that $q(\omega)$ assumes at the given flat metric $h_{ab}$,
and for which
$$\hat{\cal A}_S\circ \Psi (\gamma ) = [{\cal A}_S(h) + O({l_P\over L})]
\cdot \Psi (\gamma),\eqno(10)$$
where ${\cal A}_S(h)$ is the area of $S$ measured by the flat metric
$h_{ab}$. It turns out that such states do exist but they display a
discrete structure of a specific type at the Planck length. Roughly,
the situation is as follows. To recover $h_{ab}$, the state must have
excitations along loops which are separated by the Planck length as
measured by $h_{ab}$. If the loop separation is large, the metric
represented by the state goes to zero. In the continuum limit, on
the other hand, where the loop separation goes to zero, the metric
represented by the state diverges. It is precisely when the loop
separation is $l_P$ that one recovers $h_{ab}$. I should emphasize,
however, that this is a simplified, qualitative picture and the condition
on the loop separation is only necessary and not sufficient. The detailed
calculation is quite involved and uses the ``uniformity'' of the flat
metric $h_{ab}$. In particular, although qualitative ideas are available,
detailed constructions of weaves approximating general 3-metrics have
not yet been attempted.

That something unusual should happen at the Planck length has been
anticipated for a long time. Indeed, a number of approaches to quantum
gravity {\it begin} by assuming some discrete structure at the Planck
scale and then attempt to recover familiar macroscopic physics from it.
The situation in the present treatment is quite different. Here, one
begins with well-established principles of general relativity and quantum
mechanics and combines them using loop variables. The framework then
{\it predicts} that there should be a discrete structure, and indeed of
a specific type, at the Planck scale.

Note finally that although I have used the loop representation here
for concreteness, the same result could have been obtained using the
connection representation where quantum states are holomorphic
functionals $\Psi(A)$ of the connection $A_a^i$. It is the use of loop
variables in the regularization process that is critical to the argument.
\medskip

{\sl 4.2 Knots and links.}

In the canonical approach to quantum general relativity, dynamics is
governed by constraints. Consequently, a key step in the program is to
impose quantum constraints to single out physical states. It is here that
the simplicity of the expression (1) plays an important role: In sharp
contrast to the Wheeler-DeWitt equation of quantum geometrodynamics,
infinitely many solutions to the full set of quantum constraints has
been available in quantum connection-dynamics.

Almost all known solutions have been obtained using the loop
representation. Furthermore, we have not been able to translate the
answer back into the connection (or the metric) representation in most
cases. Thus, while the loop representation is not essential in the
discussion of kinematic issues of section 4.1, it {\it does} seem to play
a fundamental role in the discussion of dynamics.

In broad terms, so far, two avenues have been followed to solve the
constraints. The first, initiated by Rovelli and Smolin [6] and
adopted with some variations by Blencowe [22] and others [23],
deals directly with suitable functionals of loops. The second
approach, initiated by Gambini and his collaborators [24], begins by
introducing what one may call ``$SU(2)$-form factors'' of loops. That is,
given any loop, one introduces certain fields on the spatial 3-manifold
which capture precisely that information about the loop which needed to
construct traces of holonomies of arbitrary $SU(2)$ connections. (These
are non-trivial generalizations of the ``$U(1)$-form factors''
introduced in [17] to discuss quantum Maxwell fields in the loop
representation.) Quantum states are then represented as suitable
functionals of the form factors. An advantage of this approach is
that the states have a form which is familiar from other field
theories: they are functionals of fields on 3-manifold rather than
on the loop space. In both approaches, one formulates the quantum
constraints as linear operators on the wave functionals one is dealing
with and looks for their kernel. The detailed techniques used are,
however, different and the two approaches complement each other rather
well.

The results may be summarized as follows. A key feature of the loop
representation is that the Gauss constraint is automatically satisfied;
everything one does is manifestly gauge invariant. It therefore remains
to impose the vector and the scalar constraints. The vector constraint
implies that the physical states are functionals only of diffeomorphism
equivalence classes of loops, i.e., of knot classes of individual smooth
loops, or link classes of smooth multi-loops. In the Gambini approach,
these knot invariants are constructed out of the ``form factors''; one
is led to use (infinite-dimensional) differential geometrical methods to
construct these invariants. For example, the Gauss linking number arises
from a metric on the infinite dimensional space of divergence free vector
fields which constitute the simplest of the ``form factors''. This is
an interesting development in mathematical physics, and may play a role
also outside quantum gravity. In general, however, the states may have
support on loops with self-intersections or corners. One is therefore led
to consider ``generalized'' knot and link classes. Thus, any functional on
the space of loops (possibly with self-intersections and corners)
which assumes the same value on loops related by smooth
diffeomorphisms satisfies the vector constraint. Futhermore, this is the
{\it general} solution! This result is probably the most
striking and definitive application of the loop representation to
quantization of diffeomorphism invariant theories. The fact that the space
of knot (or link) classes --now the domain space of quantum states-- is
discrete is expected to simplify a number of mathematical problems in the
remainder of the program.

The full meaning of the quantum scalar constraint, on the other hand, is
still quite unclear. Therefore, although a number of strategies have been
successfully employed
to find an infinite dimensional space of solutions, we have no handle on
how complete this space is. Rovelli and Smolin [6] showed that any loop
functional which has support only on smooth loops without self-intersection
solves the scalar constraint, whence a functional of knot (or link) classes
of smooth loops satisfies {\it all} constraints. Blencowe [22] obtained some
additional solutions which are topological in origin; they are associated
with flat connections (on the spatial 3-manifold) which cannot be gauged
away. Br\"ugmann, Gambini and Pullin [25] have used Gambini's framework
[24] to show that certain knot polynomials also solve all constraints.
These developments represent definite progress especially since {\it no}
solution is known to the Wheeler-DeWitt equation in the full, non-truncated
quantum geometrodynamics. However, a number of issues still remains
unresolved. First, the physical significance of these solutions is still
quite obscure. Furthermore, it is not even clear if they will have a finite
norm with respect to the correct inner-product that, e.g., implements the
reality conditions
\footnote{The requirement that the norm be finite can in fact carry
the crucial physical information. For example, if one solves the
eigenvalue equation for the harmonic oscillator in the position
representation, one finds that there exist eigenstates $\Psi(x)$ for
{\it any} value of energy. It is the requirement that the norm be
finite that enforces both positivity and quantization of energy. Note
however that whether all solutions to the eigenvalue equation are
normalizable depends on the choice of representation. For example, if
states are taken to be holomorphic functions of $z=q-ip$, every eigenstate
$\Psi(z)$ turns out to be normalizable whence the conclusion that energy
is positive and quantized can be arrived at simply by solving the
eigenvalue equation. Of course, a priori it is not clear whether the
loop representation is analogous to the $x$ representation of the $z$
representation in this respect.}
Finally, the regularization procedure used for finding these
solutions is not as clear-cut as that discussed in section 4.1.
\medskip

{\sl 4.3 Global picture}

The results discussed in the previous two sections are based on the
general framework introduced by Rovelli and Smolin [6] in quantum
general relativity. However, at certain steps, their construction is
only formal. It is therefore appropriate to ask if there is a precise
mathematical structure underlying the loop representation and if one
can use the loop space methods in familiar theories where we already
knows what the``correct answer'' is.

The gain confidence as well as physical insight into these methods,
therefore, (source-free) Maxwell theory and linearized gravity were
considered in detail. In the Maxwell case [17], the starting point is
the Bargmann representation in which states are holomorphic functionals
of positive frequency Maxwell fields and one arrives at the loop
representation by performing a transform along the lines suggested by
Rovelli and Smolin in [6]. However, now one can show that this loop
transform exists rigorously. Since in full general relativity there is
no notion of positive and negative frequency decomposition, in the
case of linearized gravity we used variables which are the
analogs of the $A_a^i$ and $E^a_i$ used in the full theory. Quantization
was completed in the connection representation (see chapter 11 in [1]
and also [16]) as well as the loop representation [18] following the
general program outlined in section 2. In particular, the correct
inner-product on the physical states is recovered without having to
make explicit use of the underlying Poincar\'e group. However, now,
certain new features are encountered. In the connection representation,
while the quantum states are holomorphic functionals of the (self dual,
linearized) connections of one helicity as anticipated, they are
``holomorphic distributions'' of the (self dual, linearized) connections
of the other helicity. While this causes no problem in the connection
representation itself, now the loop transform fails to be well-defined.
The loop representation is constructed directly [18] starting from the
linearized loop variables following a second strategy suggested by Rovelli
and Smolin [6]. However, we find that states are now functionals of
``thickened'' loops. Somewhat surprisingly, for any (non-zero) value of the
``thickening parameter'', the resulting loop representation is isomorphic
to the Fock representation (where, however, the isomorphism does depend
on the value of the parameter). Thus, both the connection and the loop
representations based on self-dual $A_a^i$ and real $E^a_i$ are
somewhat different from those based on negative frequency $A_a^i$
and positive frequency $E^a_i$. To summarize, it {\it is} reassuring that
the general ideas underlying loop quantization successfully reproduce
known physics in the case of the Maxwell theory and linearized gravity.
However, when looked at in detail, one finds that the techniques used here
do differ from those employed in full quantum general relativity.
Furthermore, because of the absence of a background metric, it does not
seem possible to change the strategy and use, e.g., thickened loops in the
full theory.

Another development [7] addresses the issue of precision. The goal here
was to sharpen some of the heuristic notions used by Rovelli and
Smolin [6] by constructing a precise algebra of loop operators and a proper
representation theory for it. This was achieved by using the Gel'fand
spectral theory of $C^\star$-algebras (see e.g. [26]), the
$C^\star$-algebra in question being built directly out of quantum
holonomy operators $\hat{Q}_\gamma$. These are the quantum analogs of the
holonomy functionals $Q_\gamma$ of Eq.(3) where, however, {\it no assumption
is made on the existence of the operator-valued distribution}
$\hat{A}_a^i(x)$. Thus, $\hat{Q}_\gamma$ are the primary objects; in the
quantum theory, $\hat{A}_a^i$ may not even exist. The representation theory
leads us directly to the connection representation. However, the quantum
states are functionals not on the space {\sl A/G} of smooth connections
modulo gauge transformations but on a space $\Delta$ which is a
``completion'' of {\sl A/G}. This is analogous to the quantum theory of
scalar fields where the states are functionals not on the space of scalar
fields but on a completion thereof containing {\it distributions}.
Indeed, the structures involved in the present case closely parallel those
involved in the case of a scalar field. Finally, in this framework, the
loop transform is {\it rigorously} defined and hence one has fuller control
on and a deeper understanding of the mathematical structures involved.
Unfortunately, however, as it stands the framework is {\it not} directly
applicable to quantum general relativity. The problem is with the
non-triviality of the Lorentzian reality conditions. Had the connection
been real --as in Yang-Mills or topological theories, or Euclidean general
relativity-- everything would have been straightforward. With the
Lorentzian reality conditions, however, the $\star$-relations in the
required $C^\star$-algebra are hard to introduce whence, as the matters
stand, one cannot even get started. To summarize, once again, these
results give us confidence in the general ideas surrounding the loop
representation. They have proven to be especially useful in certain
topological field theories, including 2+1-dimensional general relativity,
where they provide a unifying and precise mathematical framework. However,
a significant new input is needed to make them directly applicable to
full quantum general relativity. As of now, the treatment of the
3+1-dimensional theory remains heuristic.
\bigskip

{\bf 5.Discussion}

In the last two sections, I sketched the new developments that have
occurred in the quantization program since the Goa conference. In this
section, I will summarize the open problems and directions of current
research.

At this stage of the program, I believe that the major mathematical
problems are the following: i)Understanding the structure of the space
of physical states found so far and addressing the issue of completeness;
ii)Interpreting operators that act on the physical
states, e.g., by finding their classical analogs. These would be the
``Dirac observables'', i.e. functions on the phase space which weakly
commute with constraints; and, iii)Finding ways to impose the reality
conditions to select the inner product on physical states. These
mathematical issues are related to the key physical questions
facing the program: i)Resolving the issue of time; ii)Developing
approximation methods to help with the interpretation of states and
observables; iii)Dealing squarely with the Planck regime where the
approximations break down and new concepts are needed. In particular,
one would have to learn how to interpret quantum mechanics in absence
of a background space-time geometry.

There is, of course, no a guarantee that these steps can be
satisfactorily completed. Yet, there do exist a number of model systems
in which the mathematical problems have been fully resolved and
significant progress has been made on the physical issues. These models
have been obtained by truncating general relativity in various ways and
therefore share several key features with the full theory. I will first
outline these developments and then indicate the strategies they suggest
to address the problems listed above.

The first model is obtained by a weak-field truncation of general
relativity [27]. The idea is to choose a background point in the
phase space, $A_a^i = 0$ and $E^a_i = h^a_i (\equiv {\rm flat})$,
corresponding to flat space-time, consider the deviations $\Delta A_a^i
= A_a^i -0$ and $\Delta E^a_i= E^a_i -h^a_i$ and keep in the expressions
of operator constraints terms which are at most quadratic in these
deviations. These truncated constraints are then imposed on wave functions
$\Psi [A]$ in the connection representation. To see the result, let us
first use the (c-number) flat background triad $h^a_i$ to convert internal
indices to space indices and define operators $\Delta\hat{A}_{ab}$ and
$\Delta\hat{E}_{ab}$. Next, let us decompose these operators into
anti-symmetric and symmetric parts and the symmetric parts into trace,
longitudinal and transverse-traceless parts. The result of imposition of the
constraints is then as follows. The Gauss constraint is essentially a
functional differential equation which tells us how a physical state $\Psi
(A)$ changes when we change $A_a^i$ by a purely anti-symmetric term.
Similarly, the vector constraint provides a functional differential equation
determining the dependence of $\Psi(A)$ on the longitudinal part of
$A_a^i$ while the scalar constraint determines the dependence on the
trace-part. Thus, the dependence of the wave function on the (symmetric)
transverse, traceless part is arbitrary (although it does of course have
to be holomorphic) and the constraints determine its dependence on
the remaining parts of $A_a^i$. Thus, as expected, the constraints ensure
that the true degrees of freedom lie in the transverse trace-less part of
$A_a^i$.

However, this is not all: the specific form of the quantum constraints
contains more information. Let us focus on the scalar constraint.
Its precise form can be reduced to:
$$i\hbar {\delta \over \delta A^T_{\rm im}(\vec x)} \Psi (A) = ({}^{\rm
   STT}\!\hat{A}^\star_{ab}(\vec x)\>\- {}^{\rm STT}\!\hat{A}^{ab}
    (\vec x))\circ \Psi (A)\> ,\eqno(11)$$
where $A^T$ and $A^{STT}_{ab}$ denote the trace and the symmetric,
transverse, traceless parts of $A_a^i(\vec x)$, the subscript ``${\rm im}$''
stands for ``imaginary part'' and $\star$ denotes the Hermitian conjugate.
The operator on right is precisely the Hamiltonian density in the weak
field limit. Therefore, the equation can be re-interpreted as a
(bubble-)time evolution equation, where the role of ``local'' time is
played by the imaginary part of $A^T (\vec x)$. If we integrate this
equation on $\real^3$, we obtain the Schr\"odinger evolution equation.
Thus, in the truncated model, one {\it can} single out, from various
components of the connection, an appropriate time variable. When this is
done, the familiar Schr\"odinger evolution emerges simply as one of the
``components'' of the quantum scalar constraint. Note the dual role of
constraints. On the one hand, they simply constrain the form of the
allowable wave functionals: One begins with arbitrary holomorphic
functionals $\Psi (A)$ and finds that their dependence on the
anti-symmetric part, the longitudinal part and the trace is completely
fixed by their dependence on the transverse trace-less part of $A_a^i$.
In this picture, nothing ``happens''. Physical states are simply certain
type of functionals on the connection superspace. On the other hand, if
we now foliate the infinite dimensional connection superspace by level
surfaces of $A^T$, we discover that the physical states ``evolve''
from one level surface to another in a specific way and that the
operator governing this evolution is simply the Hamiltonian density.
If one adds matter sources, the right hand side of Eq.(11) is augmented
precisely by the Hamiltonian density of matter fields. Thus, in the
connection representation, we see in detail how familiar Minkowskian
physics can emerge, in the appropriate approximation, from the constraint
equations of non-perturbative quantum general relativity. Finally, note
that such a ``deparametrization'' of the truncated quantum theory is not
possible in the metric representation of geometrodynamics [28].

The remaining truncated models which have been studied in detail have
only a finite number of degrees of freedom: type I and II Bianchi models
[29], spherically symmetric gravitational fields [30] and 2+1 dimensional
general relativity [1,19,31]. However, they capture some of the genuinely
non-linear features of full general relativity. In all these cases, the
quantization program was completed in the connection representation:
there is a complete set of solutions to all quantum constraints, a
complete set of operators corresponding to the classical Dirac observables
is known explicitly, the reality conditions are well-understood and
an inner product can be introduced on physical states unambiguously.
Furthermore, in the Bianchi models and 2+1 gravity [31], the theory
can be ``deparametrized'' and Schr\"odinger evolution can be recovered
naturally
\footnote{The work on spherically symmetric models is still
unpublished; I have not seen proofs and am not sure of the status of the
issue of time.}.
In the case of 2+1 gravity, the program is completed also in the loop
representation. Now, the (2-dimensional) diffeomorphism invariance
requires that physical states can only depend on homotopy classes of
closed loops and the inner product is introduced by exploiting the fact
that the space of homotopy classes  --like the space of knot classes in
the 3+1-dimensional case-- is discrete and therefore admits a natural
measure.

These results provide a number of lessons for the full 3+1 theory. First,
in all these cases, the reality conditions turn out to be rather simple
and determine the inner-product on the space of physical states. This
is done {\it prior} to deparametrization: The mathematical structure
can be introduced before the variable corresponding to time is isolated.
This is significant because, in the full theory, it is still unclear if
there exists an exact decomposition of the components of the connection
in to the ``time part'' and the ``dynamical part''. As noted above,
however, such a decomposition does exist in the weak field truncation
and it enables one to recover familiar physics. It is conceivable that,
in the full theory, the notion of time can only be introduced in an
approximate sense. This may suffice for the purpose of interpretation.
In regimes in which the approximation is good, one could interpret the
theory using space-time concepts we are used to, while in other regimes,
we may have to get accustomed to doing physics using genuinely new
concepts, perhaps suggested by the mathematical framework. On the other
hand, had the deparametrization of the theory been essential for
introduction of the inner-product, as was widely assumed in quantum
geometrodynamics, the approximate notion of time would not have sufficed:
it is hard to do mathematical physics with approximate inner-products! It
is fortunate therefore that the reality conditions provide us with a
well-defined principle to select the inner product and the potential
fuzziness in the notion of time affects the only the physical interpretation,
where one can afford to be somewhat imprecise.

The second lesson drawn from the work on 2+1-gravity concerns the
the solutions to the constraints in the connection and the loop
representations. As in the 3+1 theory, the quantum constraints are
easier to solve in connection dynamics than in geometrodynamics. However,
initially this method of solving the constraints was criticized on the
grounds that it did not provide insight into the nature of quantum geometry.
If one can not address questions about geometry, why are these solutions
of any use? Why are they physically interesting? Isn't the ease of solving
quantum constraints an illusion? These concerns are quite legitimate.
However, they can in fact be addressed fully. For this, it is convenient
to present an analogy. Consider the problem of finding the spectrum of
the hydrogen atom. It is now well-known that the group theoretic methods
based on the $SO(4)$-symmetry are well-tailored to solve this problem.
They tell us that the eigenstates of the Hamiltonian can be labelled by
kets $\ket{n,l,m}$, and that the eigenvalues of the Hamiltonian, total
angular momentum and the z-component of angular momentum are given by
$-13.6{\rm ev}/n^2$, $l(l+1)\hbar^2$ and $m\hbar$ respectively. However,
this solution of the problem itself tells us  {\it nothing} about the
electron orbits or position or momentum distributions of electrons in
various stationary states. Coming from the standard version of the
classical theory based on position and momentum of the electron, therefore,
one might conclude at first that the $\ket{n,l,m}$-description is no solution
to the problem at all: It tells us nothing about the ``physically interesting
questions'' pertaining to the position and momentum of the electron and their
time developments. It is obvious, in the case of the Hydrogen atom, that this
criticism is ill-placed first because there is indeed very interesting quantum
theoretic information in the $\ket{n,l,m}$-description and second because,
with an appropriate change of basis, one {\it can} recover the wave
functions $\Psi_{n,l,m}({\vec x})$ from the $\ket{n,l,m}$. I would like to
suggest that the situation is rather similar in the case of 2+1 gravity.
Just as the $\ket{n,l,m}$-basis is especially well-suited to the quantum
dynamics of the Hydrogen atom, the connection and the loop basis are
well-suited to the quantum constraints of the 2+1 theory. These solutions
do carry, directly, information about certain loop-operators which
happen to be the quantum versions of classical Dirac observables, i.e.,
constants of motion. In geometrodynamics, it just happens that one is not
used to think in terms of these observables. The geometrical variables
one normally uses are analogous to the position and momenta of the
electron. Therefore, it is not surprising that one cannot extract
information about quantum geometry directly from the solutions. However,
a change of basis {\it would} enable one to extract geometrodynamical
information from the exact solutions. In fact, this transformation has been
carried out explicitly for the case when the spatial topology is that of a
2-torus  [31]. As in the case of the hydrogen atom the form of the
transformation involved is somewhat complicated.

The situation with the 3+1 theory is likely to be the same: to extract
geometrodynamical information from the solutions to quantum constraints,
substantial amount of work is needed. The situation is further complicated
by a number of factors which did not arise in the 2+1 theory. First, as
the weak field truncation shows, the loop representation is not well-suited
for extraction of time and deparametrizing the theory. In this respect, it
is rather analogous to the metric representation. In the 2+1 theory,
the loop transform is well-understood and to recover Schr\"odinger evolution,
one can go back to the connection representation which is well-suited
for deparametrization. In the full 3+1 theory, on the other hand, as we saw
in Section 4.3, the complicated nature of the reality conditions
has made it difficult to give a precise meaning to the transform,
whence, as the matter stands, we cannot transform (most of) the known
solutions to quantum constraints to the connection representation. Finally,
unlike in the 2+1 theory, the classical Dirac observables are not known
explicitly whence we cannot easily interpret the known solutions. A
satisfactory completion of the Newman-Rovelli program, outlined in
section 2, will be of tremendous help in addressing this issue.

I believe that there is an intermediate case whose analysis will clarify
these issues significantly: 3+1 dimensional gravitational fields with
one spatial Killing field. If the norm of the Killing field is constant
and the twist is zero, this truncated theory is completely equivalent to
2+1 gravity discussed above. If the norm is arbitrary but the twist is zero,
the theory is equivalent to 2+1 gravity interacting with a zero rest mass
scalar field obeying the wave equation. If the twist is non-zero, the
theory is equivalent to 2+1 gravity coupled to a doublet of scalar fields
constituting a non-linear sigma model. These theories are interesting
precisely because they are intermediate between the well understood
vacuum 2+1 theory and the full 3+1 dimensional theory. The reality
conditions are exactly the same as in the vacuum 2+1 theory: everything
is real. Therefore, many of the (at least apparent) obstacles faced by the
full, 3+1 program are absent. On the other hand, these are {\it field
theories} with an infinite number of degrees of freedom. In fact, as
2+1 dimensional theories with matter, they have been studied in
detail perturbatively [32] and appear to be {\it finite}! (There is
a curious gap in the literature, however, because while people working on
classical aspects of the theory are well aware of the relation between
3+1 and 2+1 theories mentioned above, those working in quantum theory [32,
33] are not. As a result, conceptually important connections have not been
made.) These models are now being studied extensively. We expect the loop
transform to exist rigorously: because all fields are real, the analysis
of [7] should be applicable with only small modifications. Therefore,
one should be able to solve the quantum constraints in the loop
representation, use the reality conditions to select an inner-product
{\it and} address the issue of time in the connection representation.
Finally, since perturbation theory {\it is} available, one should be
able to compare the exact results with those obtained perturbatively.
Note that, in the 3+1 interpretation, these theories do have gravitons
interacting non-linearly (with however only one polarization if the twist
is set to zero); the only restriction is that the fields depend only on
two spatial dimensions. In this sense, they capture a significant portion
of the dynamics of the full 3+1 theory.

To conclude, I will point out another direction along which work is in
progress: approximation methods. The idea [21] here is to analyse
perturbations of the weave states. The focus is on understanding the effect
of the intrinsic discreteness of weaves on perturbative results. Is there,
in particular, an effective cut-off which makes the {\it new} perturbation
theory automatically finite? These approximation methods would resemble
the standard perturbation theory for long wave lengths but differ both
conceptually and technically at short wave lengths. The general strategy
can be illustrated by an analogy. In classical general relativity, a
non-perturbative treatment is essential to obtain black hole solutions.
However, once the essential physics of these solutions is understood, one
can often ignore full dynamics and use perturbative calculations to obtain
physical results, provided, of course, the boundary conditions at the
horizons are correctly incorporated. For example, in astrophysical
calculations of stellar distributions around black holes, it suffices to use
Newtonian methods augmented only by the modified boundary conditions. I would
like to suggest that the situation may be similar with the weave states of
quantum gravity. As we saw in section 4.1, genuinely non-perturbative
methods are essential to arrive at these states. However, once their
structure is well-understood, one should be able to extract non-trivial
physical information by studying perturbations around these states,
possibly ignoring full dynamics but paying careful attention to the
inherent discreteness.
\bigskip

{\bf Acknowledgements}

Ideas and results presented in this report have come from a large number
of colleagues. It is a pleasure to thank especially Lee Smolin, Carlo
Rovelli and Ranjeet Tate for collaboration and for innumerable stimulating
discussions. This work was supported in part by the NSF grants PHY90-16733
and INT88-15209, travel funds provided by UNDP under their TOTKEN program
and by research funds provided by Syracuse University.
\bigskip
\bigskip
{\bf References}

\item{[1]} A. Ashtekar, {\it Non-perturbative canonical quantum gravity}
(Notes prepared in collaboration with R.S. Tate) (World Scientific,
Singapore, 1991).

\item{[2]}C. Rovelli, Class. \& Quant. Grav. {\bf 8}, 1613 (1991).

\item{[3]} L. Smolin,  {\it Recent developments in
nonperturbative quantum gravity}  Syracuse pre-print SU-GP-92/2-2.

\item{[4]} A. Ashtekar, J.D. Romano and R. S. Tate, Phys. Rev. {\bf
D40}, 2572 (1989).

\item{[5]} T. Jacobson, Class. \& Quant. Grav. {\bf 5}, L143 (1988);
L. Bombelli and R. J. Torrence, Class. \& Quant. Grav. {\bf 7}, 1747
(1990).

\item{[6]} C. Rovelli and  L. Smolin, Phys. Rev. Lett. {\bf 61}, 1155
(1988); Nucl. Phys. {\bf B133}, 80 (1990).

\item{[7]} A. Ashtekar and C. J. Isham,  Class. \& Quant. Grav.
(in press).

\item{[8]} J. N. Goldberg, J. Lewandowski and C. Stornaiolo,
Commun. Math. Phys. (in press).

\item{[9]} R. Capovilla, J. Dell and T. Jacobson, Phys. Rev. Lett.
{\bf 63}, 2325 (1989).

\item{[10]} E. T. Newman and C. Rovelli, University of Pittsburgh
pre-prints.

\item{[11]} R. Capovilla, Nucl. Phys. {\bf B373}, 233 (1992).

\item{[12]} P. Peld\'an, G\"oteborg pre-print ITP 92-17.

\item{[13]} J. Samuel, Class. \& Quant. Grav. {\bf 5}, L123 (1988);
R. Capovilla, J. Dell and T. Jacobson, Class. \& Quant. Grav. {\bf 7},
L1 (1990); S. Koshti and N. Dadhich, Class. \& Quant. Grav. {\bf 7}, 5
(1990).

\item{[14]} C. G. Torre, Phys. Rev. {\bf D41}, 3620 (1990).

\item{[15]} P.A.M. Dirac, {\it Lectures on Quantum Mechanics}
(Yashiva University Press, New York, 1964).

\item{[16]} A. Ashtekar, C. Rovelli and L. Smolin, Geometry
and Physics, {\bf 8}, 7 (1992).

\item{[17]} A. Ashtekar and C. Rovelli, Class. \& Quant. Grav. (in press).

\item{[18]} A. Ashtekar, C. Rovelli and L. Smolin, Phys. Rev. {\bf D44},
1740 (1991).

\item{[19]} A. Ashtekar, V. Husain. C. Rovelli, J. Samuel and L. Smolin,
Class. \& Quant. Grav. {\bf 6}, L85 (1989).

\item{[20]} A. Ashtekar and R.S. Tate, Syracuse University Pre-print;
R. S. Tate, Ph.D. Dissertation (in preparation).

\item {[21]} A. Ashtekar, C. Rovelli and L. Smolin, Syracuse University
Pre-print;  and, in preparation.

\item{[22]} M. Blencowe, Nucl. Phys. {\bf B341}, 213 (1990).

\item {[23]} V. Husain, Nucl. Phys. {\bf B313}, 711 (1989); B.
Br\"ugmann and J. Pullin, Nucl. Phys. {\bf B363}, 221 (1991).

\item{[24]} R. Gambini and A. Leal, Montevideo pre-print 91.01,
R. Gambini, Phys. Lett. {\bf B255}, 180 (1991).

\item{[25]} B. Br\"ugmann, R. Gambini and J. Pullin, Phys. Rev. Lett.
{\bf 68}, 431 (1992).

\item{[26]} W. Rudin, {\it Functional Analysis} (McGraw Hill, New York
1991); chapter 11.

\item{[27]} A. Ashtekar, In:{\it Conceptual Problems of Quantum Gravity},
(Birkh\"auser, Boston 1991).

\item {[28]} K. Kucha\v r, J. Math. Phys.  {\bf 11}, 3322 (1970).

\item{[29]} G. Gonzalez and R.S. Tate (in preparation).

\item {[30]}  T. Thiemann and H.A. Kastrup, Communication at the German
Physical Society Meeting, Berlin 1992.

\item{[31]} S. Carlip, Phys. Rev. {\bf D42}, 2647 (1990); N. Manojlovi\'c
and A. Mikovi\'c, Nucl. Phys. (in press).

\item{[32]} G. Bonacina, A. Gamba and M. Martellini; pre-print
IFUM-400/FT.

\item{[33]} M. Allen, Class. \& Quant. Grav. {\bf 4}, 149 (1987).

\end